\documentclass[aps,pre,twocolumn,superscriptaddress,amsmath,showpacs]{revtex4}

\usepackage{graphicx, enumitem}

\begin{document}
\title{Absorbing states of zero-temperature Glauber dynamics in random networks}

\author{Yongjoo Baek}
\email[]{yjbaek@kaist.ac.kr}
 \affiliation {Department of Physics,
Korea Advanced Institute of Science and Technology, Daejeon
305-701, Korea}

\author{Meesoon Ha}
\email[Author to whom all correspondence should be addressed: ]{msha@chosun.ac.kr}
\affiliation{Department of Physics Education, Chosun University,
Gwangju 501-759, Korea}

\author{Hawoong Jeong}
\affiliation{Department of Physics and Institute for the
BioCentury, Korea Advanced Institute of Science and Technology,
Daejeon 305-701, Korea} \affiliation{APCTP, Pohang, Gyeongbuk
790-784, Korea}

\date{\today}

\begin{abstract}
We study zero-temperature Glauber dynamics for Ising-like spin
variable models in quenched random networks with random
zero-magnetization initial conditions. In particular, we focus on
the absorbing states of finite systems. While it has quite often
been observed that Glauber dynamics lets the system be stuck into
an absorbing state distinct from its ground state in the
thermodynamic limit, very little is known about the likelihood of
each absorbing state. In order to explore the variety of absorbing
states, we investigate the probability distribution profile of the
active link density after saturation as the system size $N$ and
$\langle k \rangle$ vary. As a result, we find that the
distribution of absorbing states can be split into two
self-averaging peaks whose positions are determined by $\langle k
\rangle$, one slightly above the ground state and the other
farther away. Moreover, we suggest that the latter peak accounts
for a non-vanishing portion of samples when $N$ goes to infinity
while $\langle k \rangle$ stays fixed. Finally, we discuss the
possible implications of our results on opinion dynamics models.
\end{abstract}



\maketitle

\section{\label{sec:intro}Introduction}

Glauber dynamics~\cite{Glauber1963} is one of the simplest ways
to implement the ordering dynamics of Ising spin systems,
which is defined such that detailed balance holds at equilibrium
described by the canonical ensemble.
When an Ising system evades analytical explanation,
we can numerically construct the equilibrium ensemble of the system
by running Glauber dynamics until the steady state is reached.
However, one may ask whether the ensemble created by Glauber dynamics
faithfully represents the equilibrium ensemble of the Ising system,
a question that needs to be addressed case by case with caution.

The answer is negative when the dynamics occurs at zero
temperature. Lack of noise and the limited range of interaction
trap the system in a limited region of phase space, or an {\em
absorbing state}, from which the ground state with parallel
neighboring spins (the only equilibrium configuration at zero
temperature) cannot be reached. For regular lattices, such
trappings occur whenever the dimension $d$ is greater than 1. For
$d = 2$, around 30\% of systems form stripes of domains and do not
evolve any longer~\cite{Spirin2001+Barros2009, Spirin2001a}. For
$d \ge 3$, an overwhelming fraction of systems fail to reach the
ground state. They instead stabilize into states of higher energy
and complicated domain shapes~\cite{Spirin2001a,Olejarz2011}.
The issue of absorbing states has been discussed for other kinds
of substrates as well, especially for complex networks in the
context of opinion dynamics~\cite{Svenson2001, Haggstrom2002,
Castellano2005, HZhou2005+2007+Castellano2006+Herrero2009,
Uchida2007}. Here, the relevant question is whether a society can
reach a consensus through the local majority rule. The simplest
and most studied case is the Erd\H{o}s--R\'{e}nyi
(ER)~\cite{Erdos1959+Gilbert1959} random network, which is
characterized by the system size $N$ and the average degree
$\langle k \rangle$. Earlier numerical
studies~\cite{Svenson2001,Castellano2005,ERComment} have shown that ER
networks fail to reach the ground state in some cases.
Combinatorial calculations by
H\"{a}ggstr\"{o}m~\cite{Haggstrom2002} proved a stronger statement
that the system always fails to reach the ground state in the
limit when $N\to\infty$ while $\langle k \rangle$ is finite.

\begin{figure}[t]
    \includegraphics[width=0.9\columnwidth]{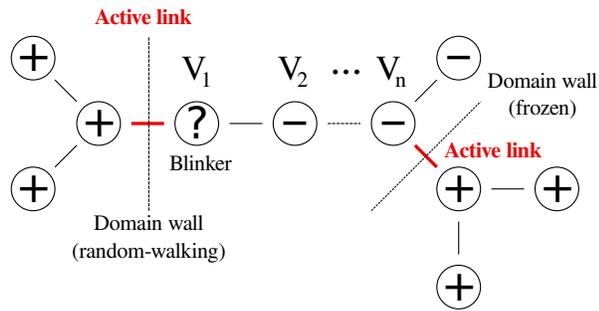}
    \caption{\label{fig:domain} (Color online)
    An example of a system in which only blinkers are flippable.
    Active links are highlighted in boldface (red) and the blinker is marked with a question mark ``?''.
    Successive flipping of spins at nodes $V_1$, $V_2$, \ldots, $V_n$ leads to the merger of two domain walls
    and further relaxation. However, the time required for the process may be long (growing like $n^2$),
    giving a false impression of saturation. This matter is addressed by the acceleration algorithm described
    in Sec.~\ref{sec:method}.}
\end{figure}

Knowing that the system fails to reach the ground state, we wonder
how close the system can get to it. The closeness of the system to
the ground state can be represented by the {\em active link
density} $l_A$, which is the fraction of links joining spins of
opposite signs as illustrated in Fig.~\ref{fig:domain}. The ground
state has $l_A = 0$, so greater $l_A$ indicates that the system is
farther away from the ground state. While $l_A$ is monotonically
decreased by Glauber dynamics at zero temperature, it does not
evolve any more once the system gets stuck into an absorbing
state. Hence, $l_A$ is a natural choice for characterizing an
absorbing state.

\begin{figure}[t]
    \includegraphics[width=0.9\columnwidth]{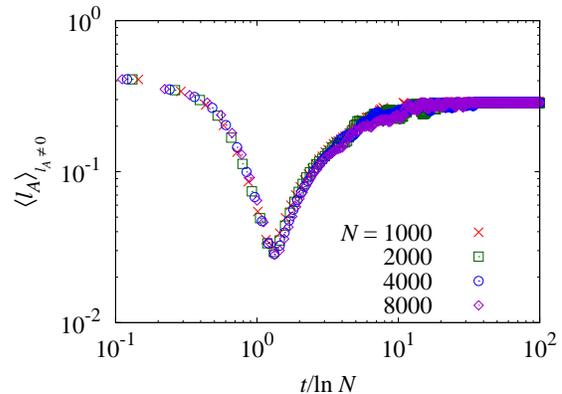}
    \caption{\label{fig:separation} (Color online)
    Dynamic scaling of $\langle l_A\rangle_{l_A\ne0}$ averaged over surviving samples
    that have never reached the ground state until time $t$. Data collapse very
    well using the scaling function $\langle
    l_A\rangle_{l_A\ne0}=f(t/\tau)$, where the relaxation time $\tau$ scales as
    $\tau\sim\ln N$.
    Numerical data are obtained from $10^5$ different network realizations
    for $\langle k\rangle=10$ as $N$ varies.
    The dip structure indicates that samples can be divided into
    two groups: the almost fully-ordered case driving the initial fall
    and the quickly absorbed case causing the subsequent rise.}
\end{figure}

H\"{a}ggstr\"{o}m~\cite{Haggstrom2002} showed that $l_A$ in
an absorbing state has a positive lower bound in the thermodynamic
limit, thus proving that full relaxation to $l_A = 0$ is
impossible. There are still more stories to be told about the
properties of $l_A$. For example, temporal evolution of $l_A$
averaged over the systems with $l_A \left(t\right) \ne 0$ hints at
the existence of at least two groups of samples with very
different behaviors of $l_A$ (see Fig.~\ref{fig:separation}, where
we provide the standard finite-size scaling analysis of the
quantity measured by Castellano {\em et
al.}~\cite{Castellano2005}). This leads us to expect that the
$l_A$ distribution has nontrivial structural features worthy of
investigation, which is still lacking.

In this paper, we present a systematic analysis of the absorbing
state distribution of $l_A$ produced by zero-temperature Glauber dynamics
in ER random networks. In particular, we focus on the influence of
parameters $N$ and $\langle k \rangle$ on the distribution. The origin of
sample-to-sample fluctuations as well as the nature of peaks are also discussed.

The paper is organized as follows. In Sec.~\ref{sec:model}, we
briefly describe zero-temperature Glauber dynamics and define
quantities of interest. Then, we describe the algorithm
proposed by Olejarz {\em et al.}~\cite{Olejarz2011} that
ensures the saturation of the system into absorbing states.
Our main results, based on extensive numerical simulations,
are presented in Sec.~\ref{sec:result}. Finally, we conclude
the paper with a summary and discussion in Sec.~\ref{sec:discussion}.

\section{\label{sec:model} Model}

\subsection{Dynamics and physical quantities}

We consider a system consisting of $N$ nodes. Each node is associated
with a spin variable $\sigma_i$ ($i = 1,\ldots,N$) whose value is either
$+1$ or $-1$. Pairs of nodes are connected by $L$ {\em randomly distributed}
links, such that $\langle k\rangle = 2L/N$ is the average degree, or
the average number of neighbors connected to each node.
Let us write $A_{ij} = 1$ if nodes $i$ and $j$ are connected,
and $A_{ij} = 0$ otherwise. An active link is a link joining nodes
with oppositely signed spins as shown in Fig.~\ref{fig:domain}.
Then, we can define the {\em active link density} $l_A$ as
\begin{equation}
    l_A = \frac{1}{L} \sum_{i<j}A_{ij}\frac{1-\sigma_ i\sigma_j}{2}.
\end{equation}

We use zero-magnetization initial states with exactly equal
numbers of $+1$ and $-1$ spins assigned at random. At each time
step, one spin is randomly chosen and updated according to
zero-temperature Glauber dynamics, Eq.~(\ref{ZTGD}). That is, a
spin flips (changes sign) with the following probability,
where $\Delta l_A$ is the change of $l_A$ resulting from the flip.
\begin{equation}
\label{ZTGD} \text{Flipping probability} = \left\{
\begin{array}{rl}
    1,      &   \text{if~} \Delta l_A < 0\\
    1/2,        &   \text{if~} \Delta l_A = 0\\
    0       &   \text{if~} \Delta l_A > 0
\end{array} \right.
\end{equation}

Zero-temperature Glauber dynamics forbids the increase of $l_A$.
Thus, spins aligned with the majority of its neighbors are ``unflippable.''
To improve the efficiency of our simulations, we employ the rejection-free
technique: instead of updating one spin among all the $N$ spins, we only
update one of the flippable spins at each time step. The Monte Carlo simulation
(MCS) time should then be increased by $1/\left(\text{number of active spins}\right)$
per update, in order for a fair comparison with the case when the rejection-free technique
is not used.

\subsection{\label{sec:method} Saturation test: absorbing states}

When zero-temperature Glauber dynamics cannot lower $l_A$ any further,
we say that the system has reached an absorbing state. If the system
has run out of flippable spins, it is certainly in an absorbing state
since the dynamics has stopped. There are, however, absorbing states
with flippable spins whose neighbors are equally split between $+1$
and $-1$. Such spins are called {\em blinkers} as they repeatedly
flip back and forth. The existence of blinkers makes the problem of
finding an absorbing state trickier.

Blinkers can mislead one to believe that the system has saturated to
an absorbing state, while $l_A$ can still be lowered as the dynamics
continues. Figure~\ref{fig:domain} illustrates this point. It shows
a system in which a single blinker is the only flippable spin at the moment.
The blinker can ``hop" randomly back and forth between nodes $V_1$ and $V_n$, which is
effectively a random domain wall motion conserving $l_A$.
If the random-walk domain wall happens to collide with the frozen domain wall
on the opposite side of the network, the system can go under further
relaxation, $l_A$ eventually going to zero. Hence, the network in
Fig.~\ref{fig:domain} has not reached an absorbing state yet. The
domain wall collision, however, takes some time (proportional to
$n^2$) due to the randomness of the domain wall motion. Until the
collision occurs, one may get a false impression that $l_A$ has
saturated and the system has reached an absorbing state. In order
to determine whether an apparent saturation of $l_A$ truly indicates an
absorbing state, we employ the algorithm proposed by Olejarz {\em
et al.}~\cite{Olejarz2011} as follows:
\begin{enumerate}
\item Until some prescribed MCS time $T$, let the system evolve
with zero-temperature Glauber dynamics. \item Apply a global
infinitesimal field, so that blinkers favor one sign over the
other and the domain walls composed of blinkers are driven in a
biased direction. The field must not be too strong, lest it should
make unflippable spins flippable or flippable spins unflippable.
The field is maintained until domain walls collide to decrease $l_A$, 
or until the system completely runs out of
flippable spins even in the presence of the field. Reverse the
global field and do the same. \item If $l_A$ never decreased in
the previous step, the system has saturated to an absorbing state
and the algorithm terminates. Otherwise, return to the previous step.
\end{enumerate}

To observe the unbiased distribution of absorbing states induced
purely by zero-temperature Glauber dynamics, one has to determine
$T$ such that no lowering of $l_A$ occurs after applying the
acceleration algorithm. This is quite a time-consuming task. So we
just accept the absorbing state found by the acceleration
algorithm even if the algorithm has lowered $l_A$. This inevitably
biases the absorbing state distribution of $l_A$, but we have made
sure that the change in the distribution is negligible if $T$ is
larger enough than $\tau$. Here, we set $T=100 \ln N$.
\begin{figure}[]
    \centering
    \includegraphics[width=0.9\columnwidth]{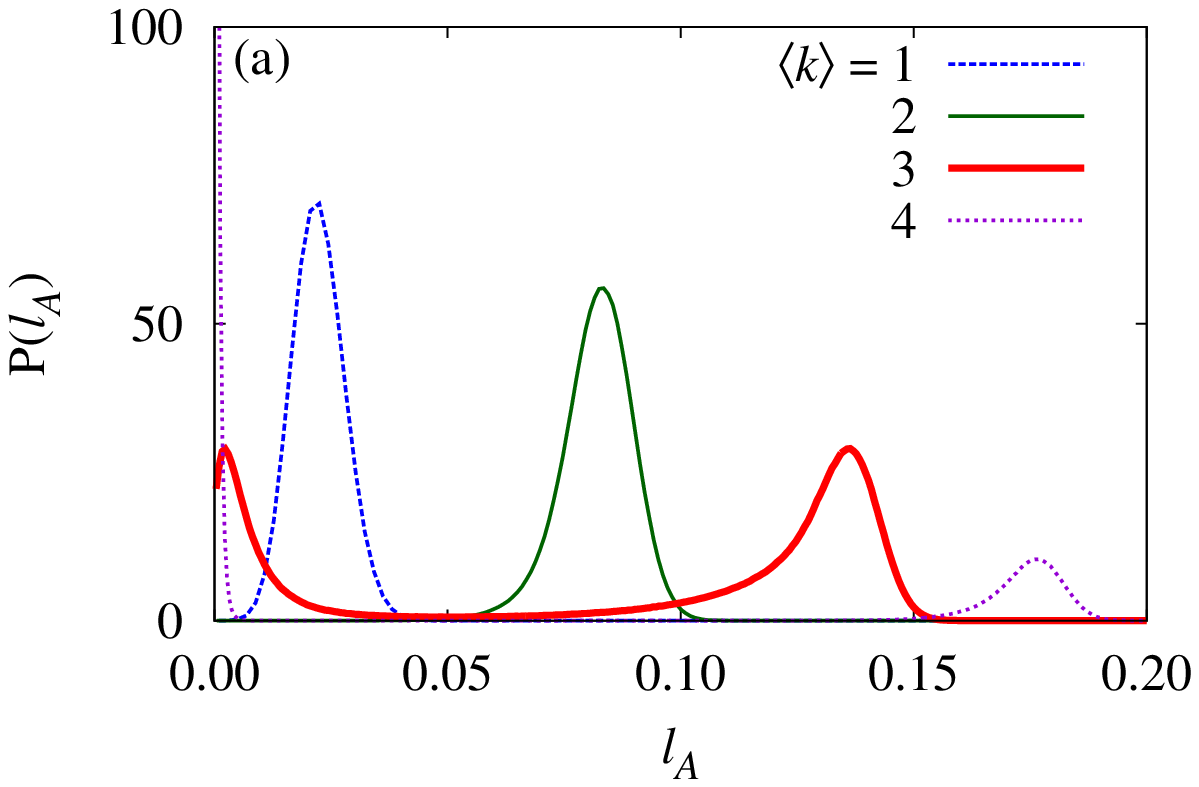} \\
    \includegraphics[width=0.9\columnwidth]{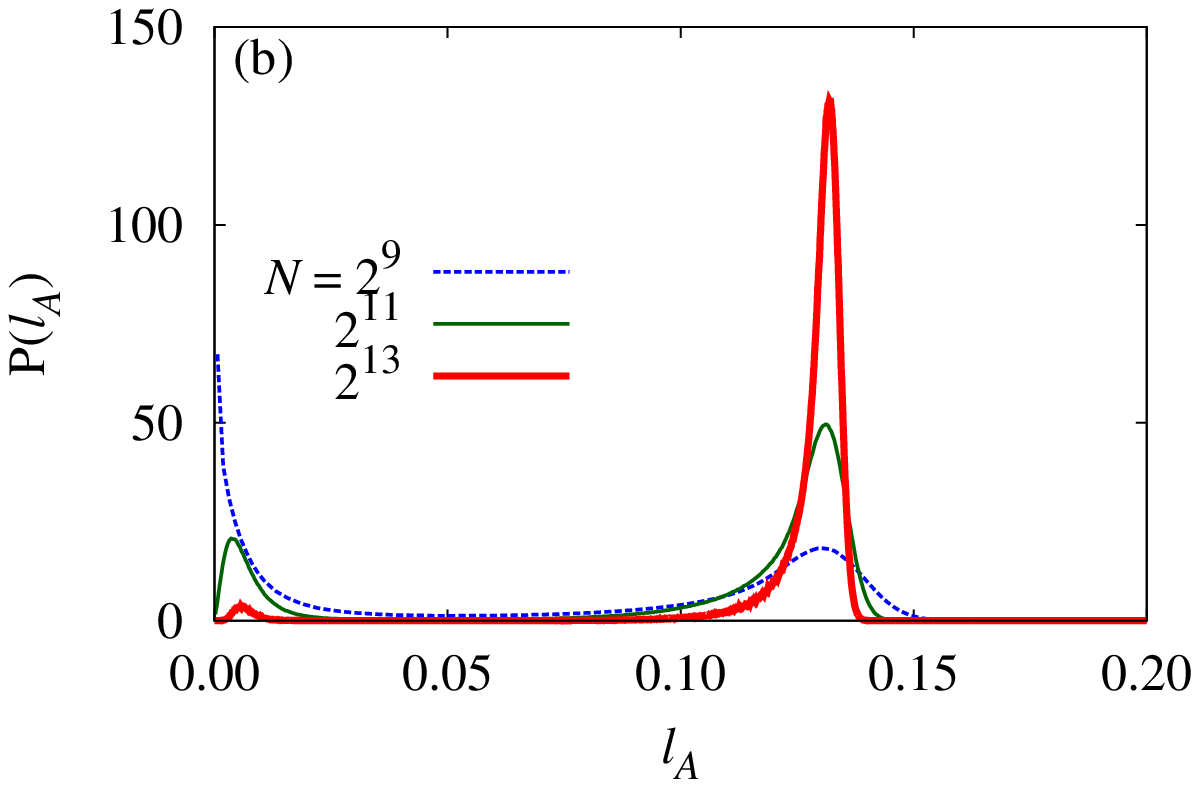}
    \caption{\label{fig:lA_dist} (Color online) Probability distribution functions
    of $l_A$ in absorbing states (a) at $\langle k\rangle=1,\ 2,\ 3,\ 4$ with
    $N=2^{10}$ and (b) at $N=2^{9},\ 2^{11},\ 2^{13}$ with $\langle k \rangle=3$.
    Numerical data are obtained from $10^2$ zero-magnetization initial configurations
    and $10^5$ network realizations. The $l_A$ distribution is essentially bimodal,
    which implies that it is a mixture of two components that need to be analyzed separately.}
\end{figure}
\begin{figure}[]
    \centering
    \includegraphics[width=0.9\columnwidth]{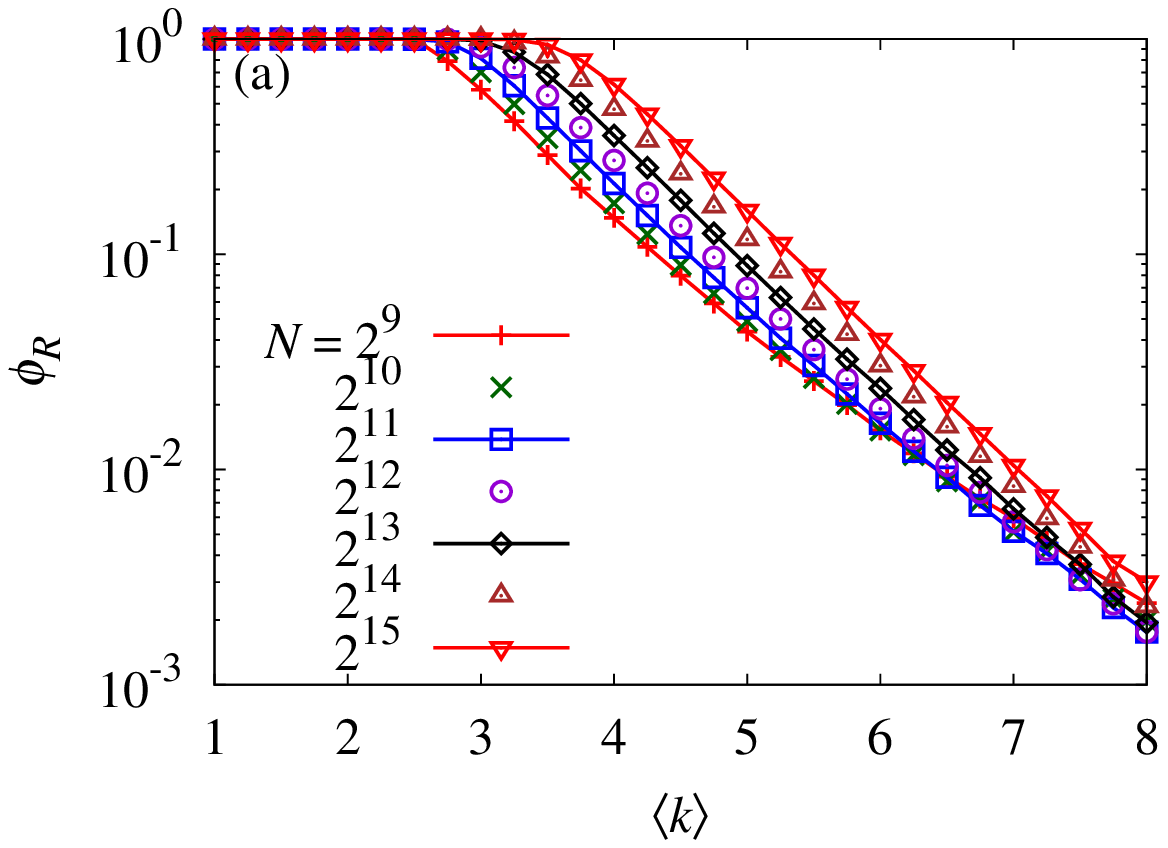} \\
    \includegraphics[width=0.9\columnwidth]{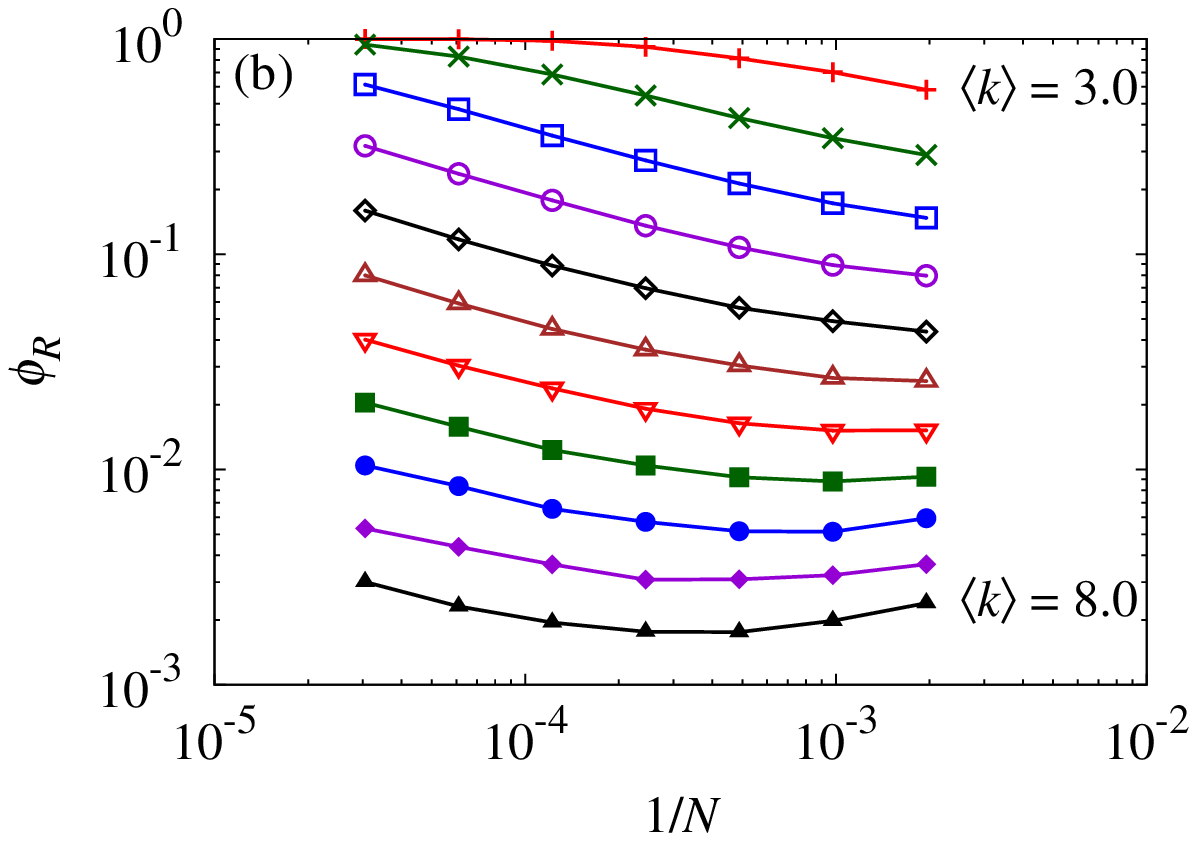}
    \caption{\label{fig:peak_mass} (Color online) The fraction of samples
    belonging to the right component, $\phi_{R}$, is calculated from
    the $l_A$ distribution.
    (a) If $N$ is fixed, the value of $\phi_{R}$ decreases with $\langle k \rangle$.
    (b) If $\langle k \rangle$ is fixed, $\phi_{R}$ eventually increases with $N$.
    Larger $N$ is required at greater $\langle k\rangle$ to observe the increasing trend.
    The lines are a guide to the eyes.}
\end{figure}

\section{\label{sec:result} Numerical results}

Utilizing the acceleration algorithm to find true absorbing states of
zero-temperature Glauber dynamics in ER networks, we obtain
the probability distribution function $P\left(l_A\right)$ of
the active link density $l_A$ in absorbing states through
extensive numerical simulations covering both different runs
on a given ER network and different network realizations.
$P\left(l_A\right)$ is typically bimodal with the left peak
near $l_A = 0$ and the right peak at a larger value of
$l_A$ (see Fig.~\ref{fig:lA_dist}).
Such a shape of the distribution is consistent with the temporal behavior of
$\langle l_A\rangle_{l_A\ne0}$ in Fig.~\ref{fig:separation}:
both imply that there are two groups of samples with very different properties.
From this perspective, we analyze $P\left(l_A\right)$ assuming that
the distribution always consists of two components, the left and the right.
$P\left(l_A\right)$ may appear unimodal in some cases, as illustrated by
the curves at $\langle k \rangle = 1$ and $\langle k \rangle = 2$
in Fig.~\ref{fig:lA_dist}(a). Even in such cases, we can assume that there is
an unobserved peak of zero (or extremely small) magnitude. An example is
shown in Fig.~\ref{fig:lA_dist}(b), where the left peak seems to shrink
indefinitely as $N$ grows, leaving only the right peak to be observed.

In the following subsections, we systematically investigate the finite-size effect
on $P\left(l_A\right)$ with the goal of obtaining the shape of the distribution
in the thermodynamic limit. We present numerical evidence that the $l_A$ distribution
consists of at most two delta peaks in the thermodynamic limit,
and that a nonvanishing proportion of samples belongs to the right peak.
Clues as to the nature of the peaks and the origin of sample-to-sample fluctuations of $l_A$
are also discussed. Finally, we provide evidence that $P\left(l_A\right)$
may have only a single delta peak far from $l_A = 0$ in the thermodynamic limit.

\subsection{Proportions of components}

We begin our analysis with the finite-size effect on the fraction
$\phi_{R}$ of samples belonging to the right component of the
$l_A$ distribution. The component corresponds to samples stuck far
away from the ground state, so higher $\phi_{R}$ indicates
greater difficulties in reaching the ground state. We are also
interested in whether the $l_A$ distribution is still bimodal in
the asymptotic limit.

Measurement of $\phi_{R}$ requires the criterion for
distinction between the two components of the $l_A$ distribution.
Without loss of generality, we choose the local minimum point
$l_{\text{min}}$ between the maximum point of the left peak
$l_{\text{max}}^{L}$ and the maximum point of the right peak
$l_{\text{max}}^{R}$ as the border between two peaks. Therefore,
$\phi_{R}$ is the fraction of samples ending up on the right
side of $l_{\text{min}}$. If there are multiple candidates for
$l_{\text{min}}$, the largest (rightmost) one is chosen for a
conservative estimation of $\phi_{R}$.
\begin{figure*}[]
    \centering
    \includegraphics[width=0.9\columnwidth]{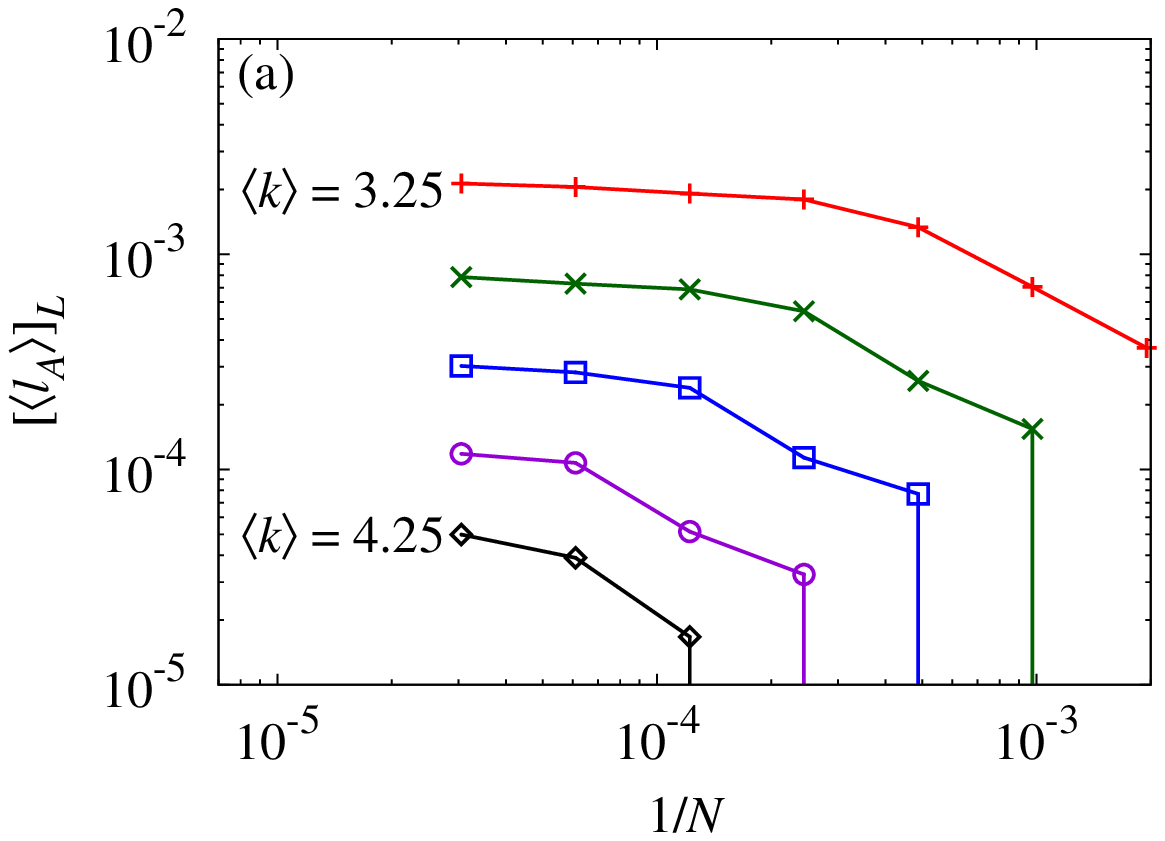} \quad
    \includegraphics[width=0.9\columnwidth]{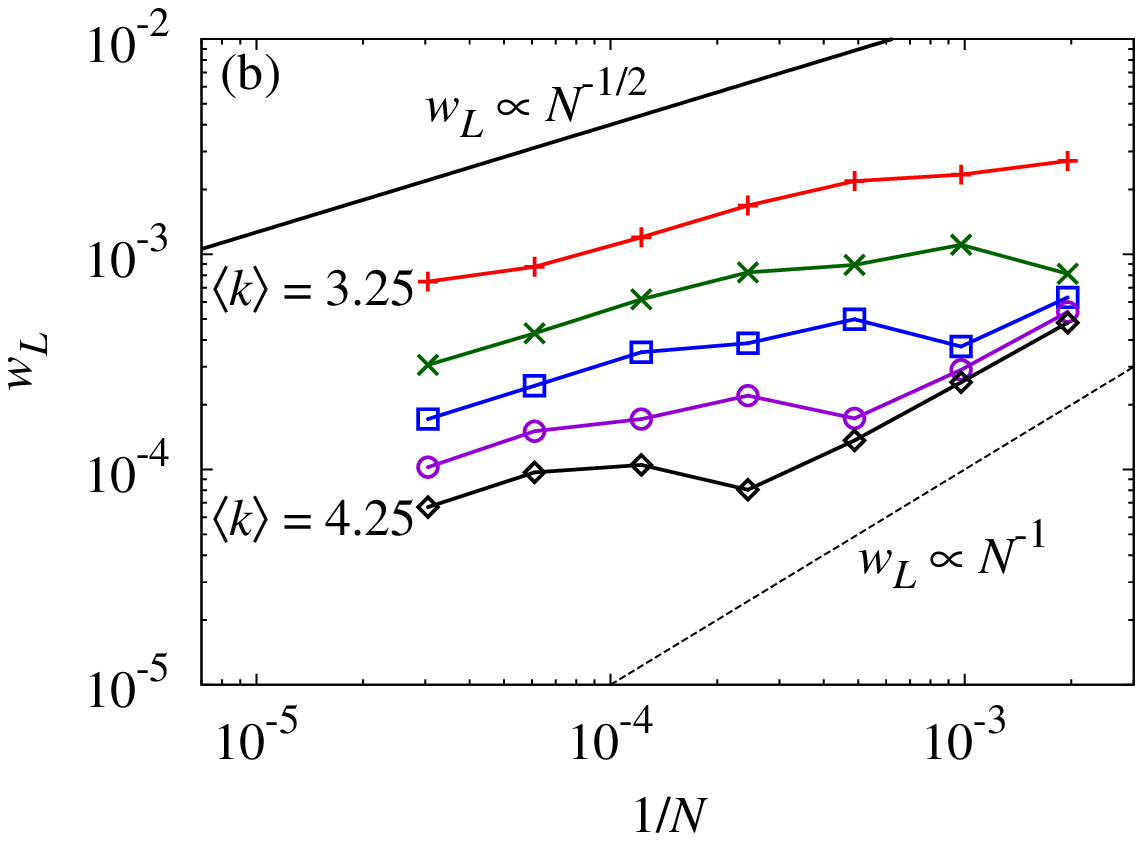} \\
    \includegraphics[width=0.9\columnwidth]{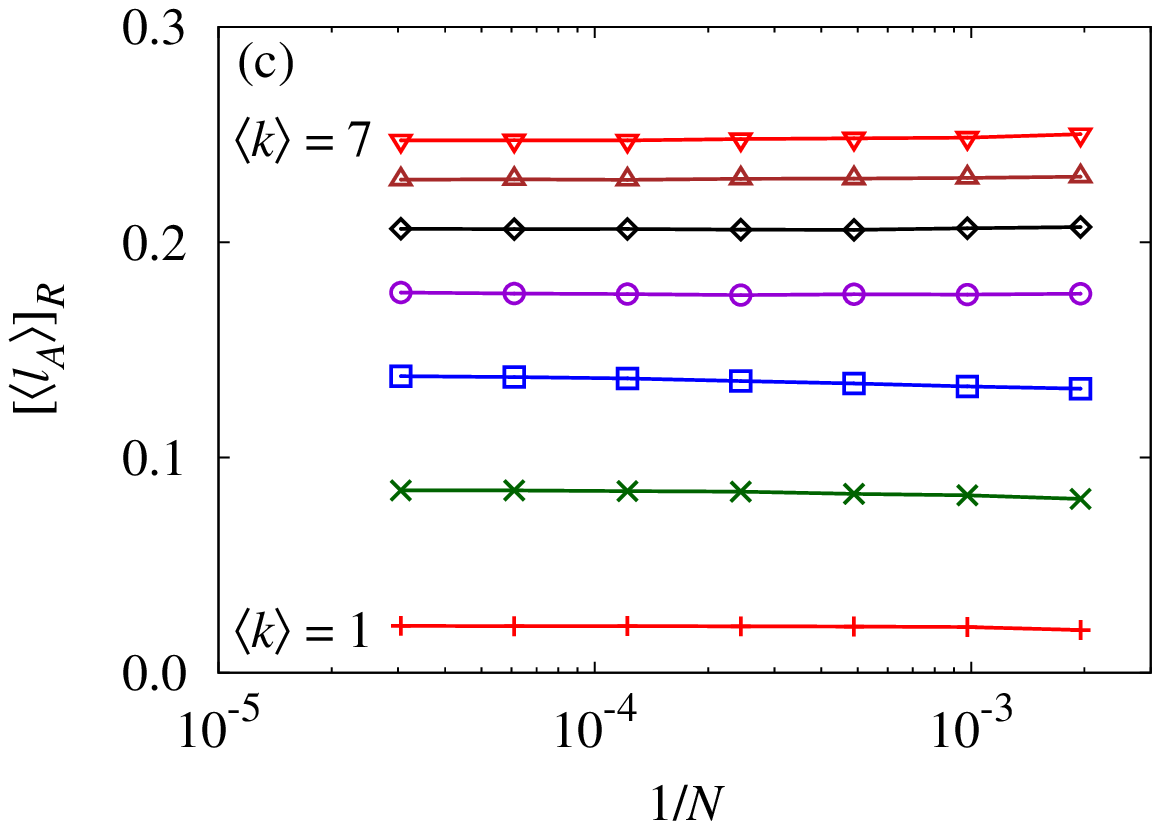} \quad
    \includegraphics[width=0.9\columnwidth]{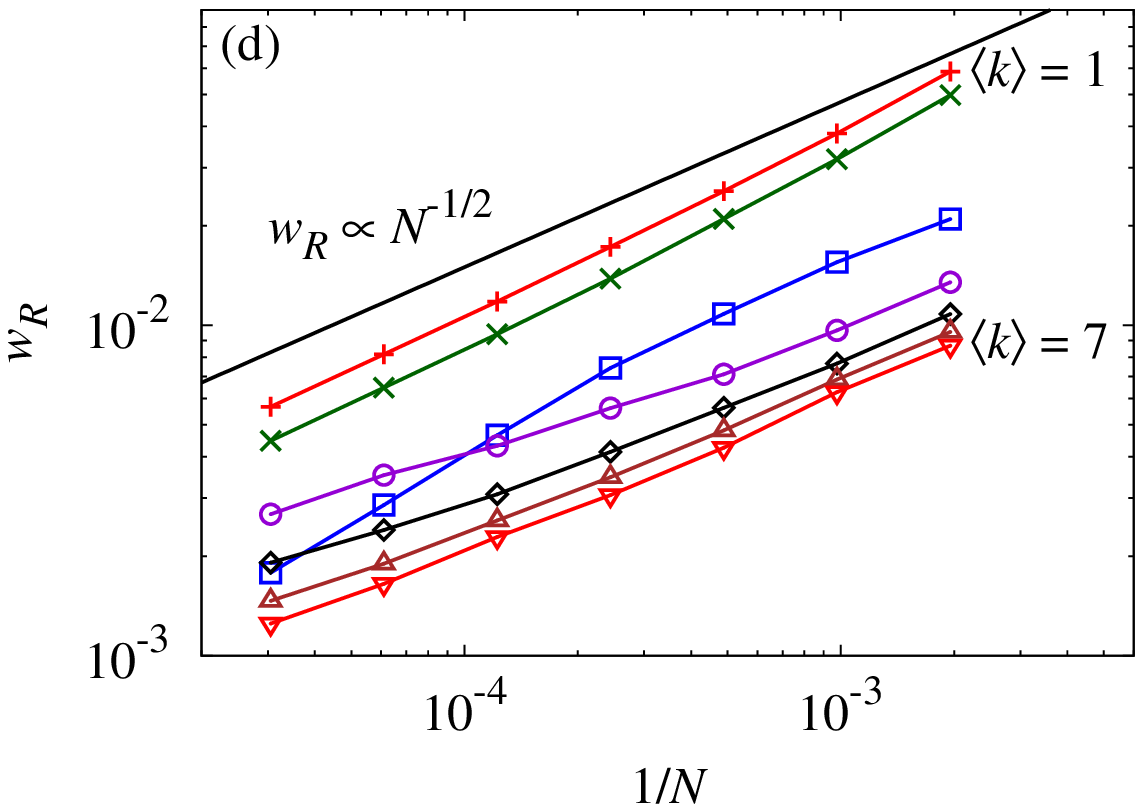}
    \caption{\label{fig:self_averaging} (Color online)
    The means and widths of two peaks of $P\left(l_A\right)$ are measured.
    (a) The left-peak mean $[\langle l_A\rangle]_{_{L}}$ is plotted at
    $\langle k \rangle=3.25,\ldots, 4.25$ increased by $0.25$ from top to bottom.
    (b) The left-peak width $w_{_{L}}$ is plotted for the same parameter sets as (a).
    (c) The right-peak mean $[\langle l_A\rangle]_{_{R}}$ is plotted for
    $\langle k \rangle=1,\ldots,7$ increased by $1$ from bottom to top.
    (d) The right-peak width $w_{_{R}}$ is plotted for the same parameter sets as (c).
    Note that the curves at $\langle k \rangle=1,\ 2$ have been vertically shifted by
    respective factors of $4$ and $5$ to avoid overlapping with other data.
    All the data are obtained from $10^3$ different initial configurations
    and $10^3$ network realizations. The lines are guide to the eyes.}
\end{figure*}

According to Fig.~\ref{fig:peak_mass}(a), $\phi_{R}$
decreases monotonically with $\langle k \rangle$ when $N$ is
fixed. This is a natural consequence of the fact that nodes are
better informed of the global situation when the connectivity is
greater. On the other hand, $\phi_{R}$ has a nonmonotonic
dependence on $N$ when $\langle k \rangle$ is fixed, as in
Fig.~\ref{fig:peak_mass}(b). $\phi_{R}$ eventually increases
with $N$, which implies that the right-peak component accounts for a
nonvanishing portion of samples in the thermodynamic limit. We
note that the crossover occurs at greater $N$ as $\langle k
\rangle$ gets larger, so observing highly connected networks of
small sizes may mislead one to conclude that the right component
becomes negligible in the thermodynamic limit. The behavior of
$\phi_{R}$ at greater $N$ shows that this is not likely to be
the case. On the contrary, one may even claim from our result that
the right peak is actually the only observed peak in the
thermodynamic limit, as can be numerically confirmed for
sufficiently small $\langle k \rangle$. We shall get back to this
bold claim later, providing more numerical evidences. For now, we
simply interpret the numerical observation for $\phi_{R}$ 
as evidence that the right component is not an artifact due to the
finite-size effect.

\subsection{Means and widths of peaks}

Now we take a closer look at the shape of $P\left(l_A\right)$, and
measure the finite-size effect on positions and widths of the two
peaks, which are defined as follows. Three extrema
($l_{\text{max}}^{L},\ l_{\text{max}}^{R},\ l_{\text{min}}$)
defined in the previous subsection divide the domain of $l_A$ into
four intervals. The medians of the intervals are denoted by $l_1 <
l_2 < l_3 < l_4$ in the ascending order. Then, the width of the
left (right) peak is defined as $w_L = l_2 - l_1$ ($w_R = l_4 -
l_3$), and the position is represented by the mean value
$\left[\langle l_A\rangle\right]_{L}$ ($\left[\langle
l_A\rangle\right]_{R}$) which is the average of $l_A$ among
samples belonging to the interval between $l_1$ and $l_2$ ($l_3$
and $l_4$). The behaviors of those peak properties are plotted in
Fig.~\ref{fig:self_averaging}. We note that both peaks are
self-averaging: the mean values show signs of saturation, and the
widths seem to decay asymptotically like $N^{-1/2}$. Hence, in
the thermodynamic limit, we claim that almost all the samples
belong to two sharp peaks whose positions are determined solely by
$\langle k \rangle$. Here follows detailed descriptions for the
analysis of each peak.

\subsubsection{Analysis of the left peak}

We present the left-peak properties for a rather limited range
$3.25 < \langle k\rangle < 4.25$, due to the following limitations
of our study. The left peak may not be observed for small values of
$\langle k\rangle$ as shown in Figs.~\ref{fig:lA_dist}(a)
and \ref{fig:peak_mass}, so its properties cannot be
measured in such cases.
On the other hand, the left peak is too sharp and too close to
zero for large values of $\langle k\rangle$, so we hit the resolution
limit of peak position and width. In such cases, the left-peak properties
show the trivial behavior of a peak of unit width
(proportional to $1/N\langle k\rangle$) located at $l_A = 0$.

Figure~\ref{fig:self_averaging}(a) shows that the value of
$[\langle l_A\rangle]_{L}$ decreases with $\langle k \rangle$,
and approaches positive limits as $N$ grows. This is consistent
with the lower bound of $l_A$ in the thermodynamic limit
derived by H\"{a}ggstr\"{o}m~\cite{Haggstrom2002}
which is positive and decreases with $\langle k \rangle$
in the parameter range observed in our simulations.
The vertical drops of $l_A$ indicate that the peak becomes
too sharp and too close to zero at smaller values of $N$, so
$[\langle l_A\rangle]_{L}$ becomes indistinguishable from zero
in those ranges.

Figure~\ref{fig:self_averaging}(b) shows that the value of $w_{L}$
has a crossover between two scaling regimes: $N^{-1}$ and
$N^{-1/2}$. As pointed out in previous paragraphs, the left peak
is so narrow for small values of $N$ that its width is just given
by the unit of $l_A$, which is inversely proportional to $N$. As
$N$ gets larger, the unit of $l_A$ decays faster than the true
peak width, revealing the $N^{-1/2}$ dependence. This crossover to
the true scaling regime occurs for the larger values of $\langle k
\rangle$ as $N$ gets larger, which is obviously related to the
shifting of vertical lines in Fig.~\ref{fig:self_averaging}(a),
implying that the two artifacts are of the same origin.

\subsubsection{Analysis of the right peak}

The right-peak properties are presented for a wider range of
parameters, as the limitations present in the case of left-peak
properties are not as severe. For the sake of clarity, we have
reduced the resolution of $\langle k\rangle$ compared to
the case of the left peak.

While Fig.~\ref{fig:self_averaging}(c) clearly shows that
$[\langle l_A\rangle]_{R}$ saturates as $N$ grows and increases
with $\langle k \rangle$, $w_{R}$ exhibits more complicated
behaviors as shown in Fig.~\ref{fig:self_averaging}(d). For
$\langle k \rangle> 3$, the curves seem to follow the $N^{-1/2}$
scaling at small $N$, but deviate from it as $N$ grows. On the
other hand, for $\langle k \rangle < 3$, the curves are
increasingly well-fitted by $N^{-1/2}$ as $N$ increases. We claim
that $w_R$ decays somewhat like $N^{-1/2}$, but deviates from it
around the regime where the left peak is just about to disappear
(or, equivalently, just about to appear). The data for $\langle k
\rangle = 3$ in Fig.~\ref{fig:self_averaging}(d) and the
distribution shapes in Fig.~\ref{fig:lA_dist}(b) give a nice
illustration of our claim. When $N$ is small, the two peaks are
clearly distinguishable, so the $N^{-1/2}$ behavior is easily
observed. As $N$ grows, the left peak dwindles and the left
component becomes indistinguishable from the right component of
the $l_A$ distribution. This makes $w_R$ appear broader than it
really is as we include part of the left component in the right
component by mistake, hence $w_R$ appears to decay slower than
$N^{-1/2}$. As $N$ grows more, the left peak becomes completely
unobservable, causing the rapid width decay and the subsequent return
to the $N^{-1/2}$ behavior. If we had observed the system at
greater $N$, the curve for $\langle k \rangle = 7$ would also have shown
the rapid width decay and the subsequent return to the $N^{-1/2}$
behavior as was shown by the curve for $\langle k \rangle = 3$.

Taking all those considerations into account, we conclude that
both components of $P\left(l_A\right)$ are self-averaging.

\subsection{Nature of the peaks}

\begin{figure*}
    \centering
    \includegraphics[width=0.9\columnwidth]{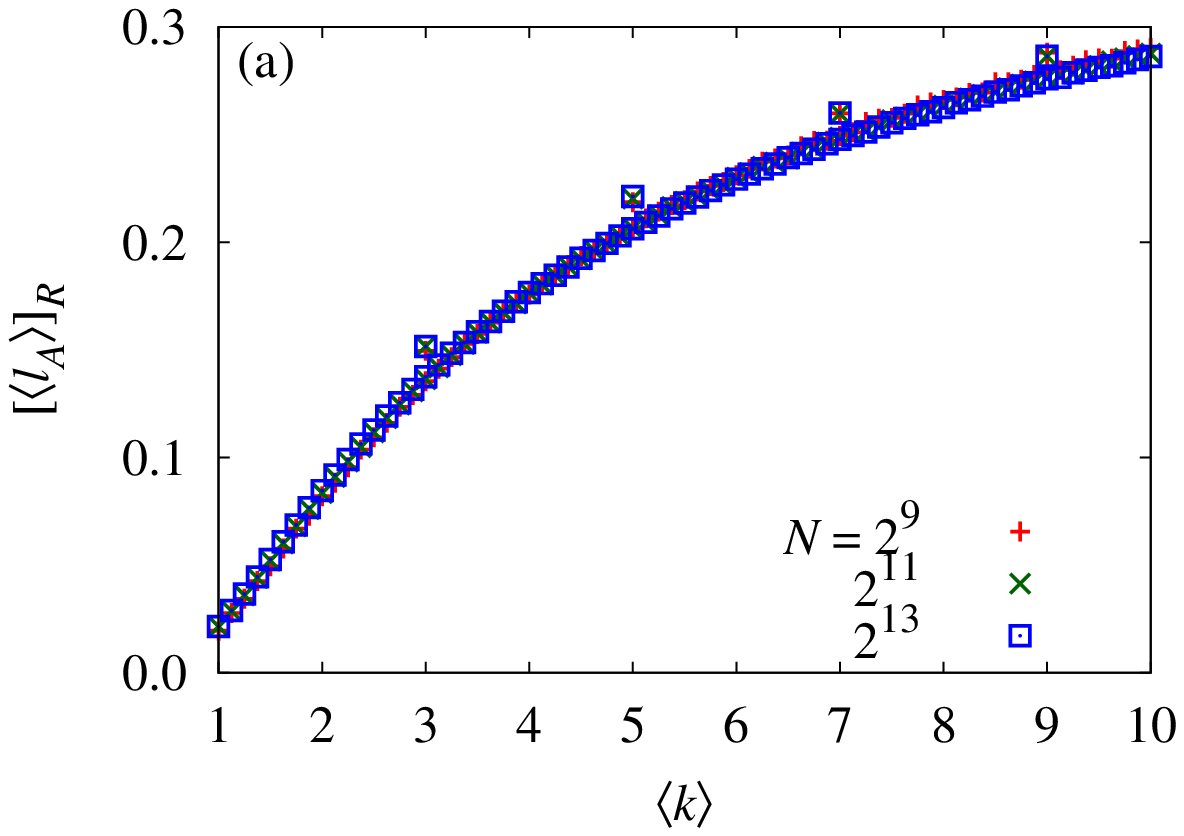}
    \includegraphics[width=0.9\columnwidth]{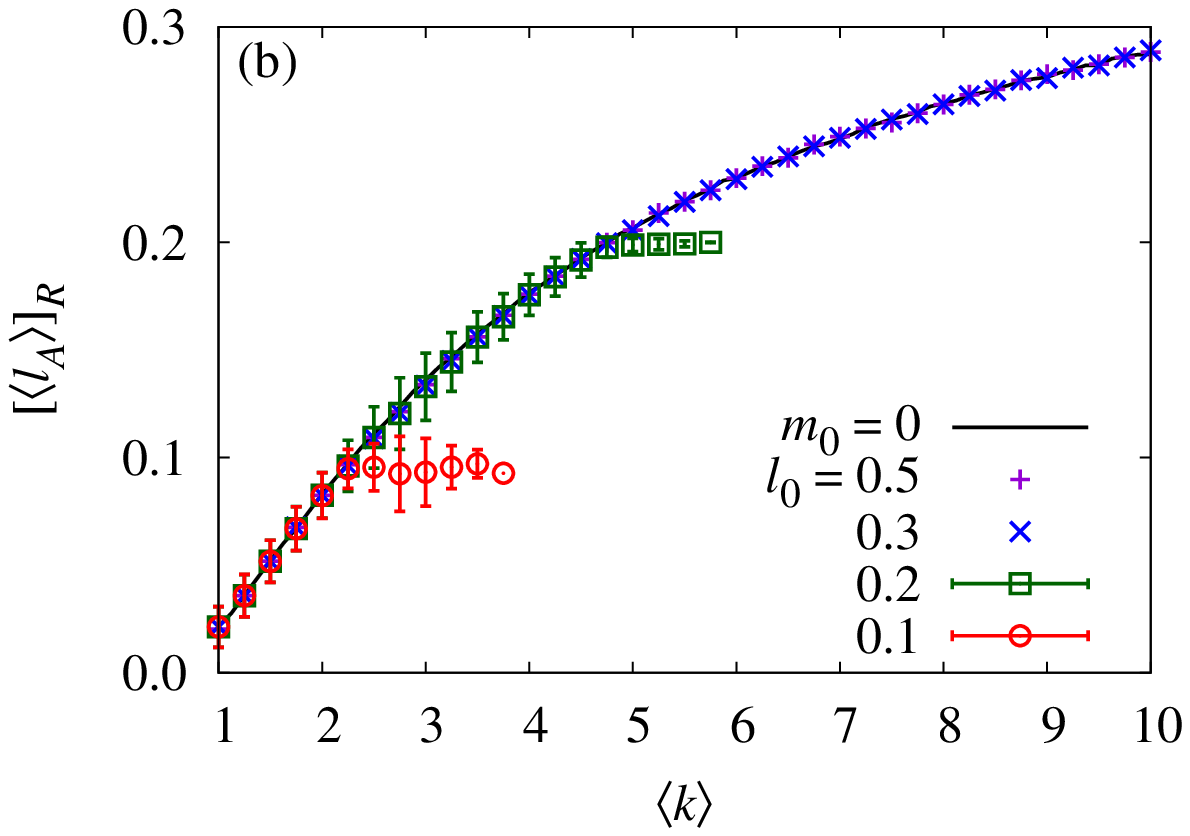}
    \caption{\label{fig:rightmean} (Color online)
    (a) The right-peak mean of $l_A$ measured for $10^7$ ER network samples (lower points)
    and $10^6$ regular random network samples (upper points).
    Lack of degree fluctuations in odd-degree regular random networks does not make much difference
    in the right-peak mean, while the right peak completely disappears in even-degree
    regular random networks.
    (b) The same quantity measured for different initial magnetization $m_0$
    or initial active link density $l_0$ in $10^6$ ER network
    samples. The result suggests that for each value of $\langle k \rangle$, the right peak corresponds
    to the only stable spin configurations with approximately zero magnetization.}
\end{figure*}
\begin{figure*}
    \centering
    \includegraphics[width=0.9\columnwidth]{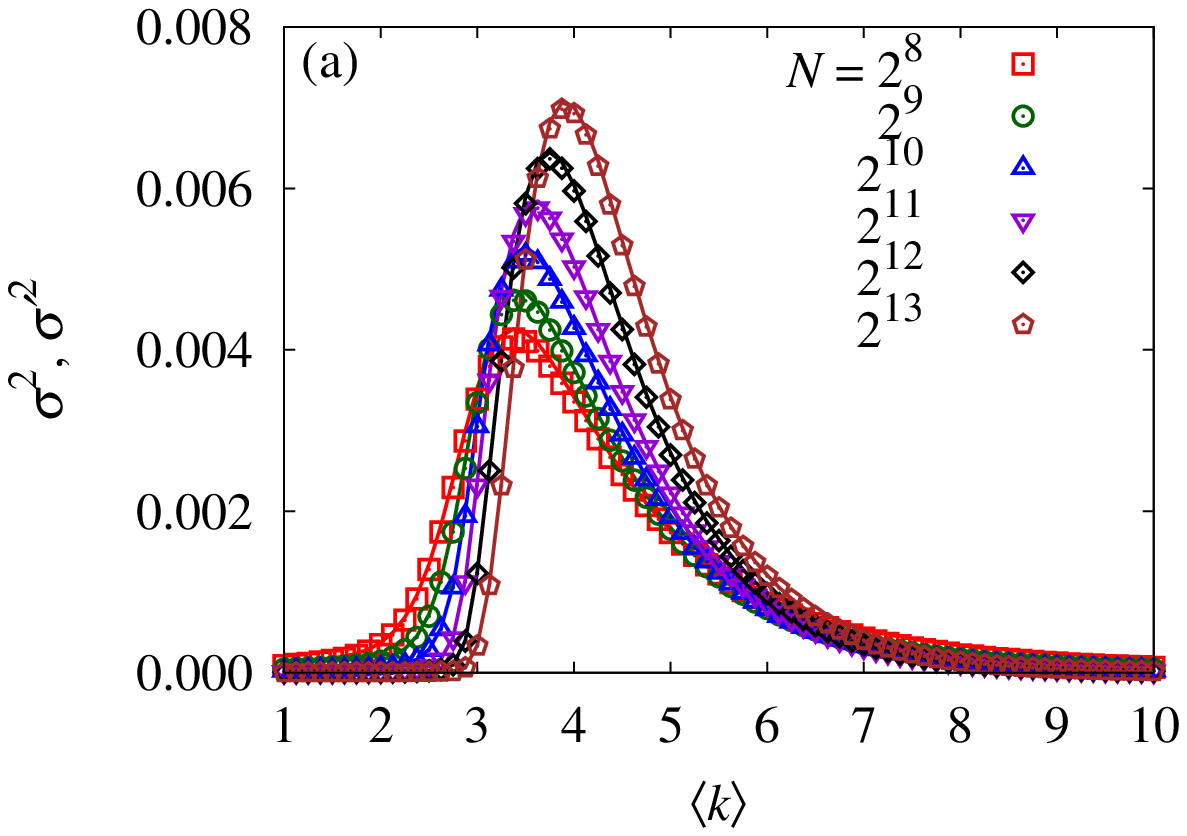} 
    \includegraphics[width=0.9\columnwidth]{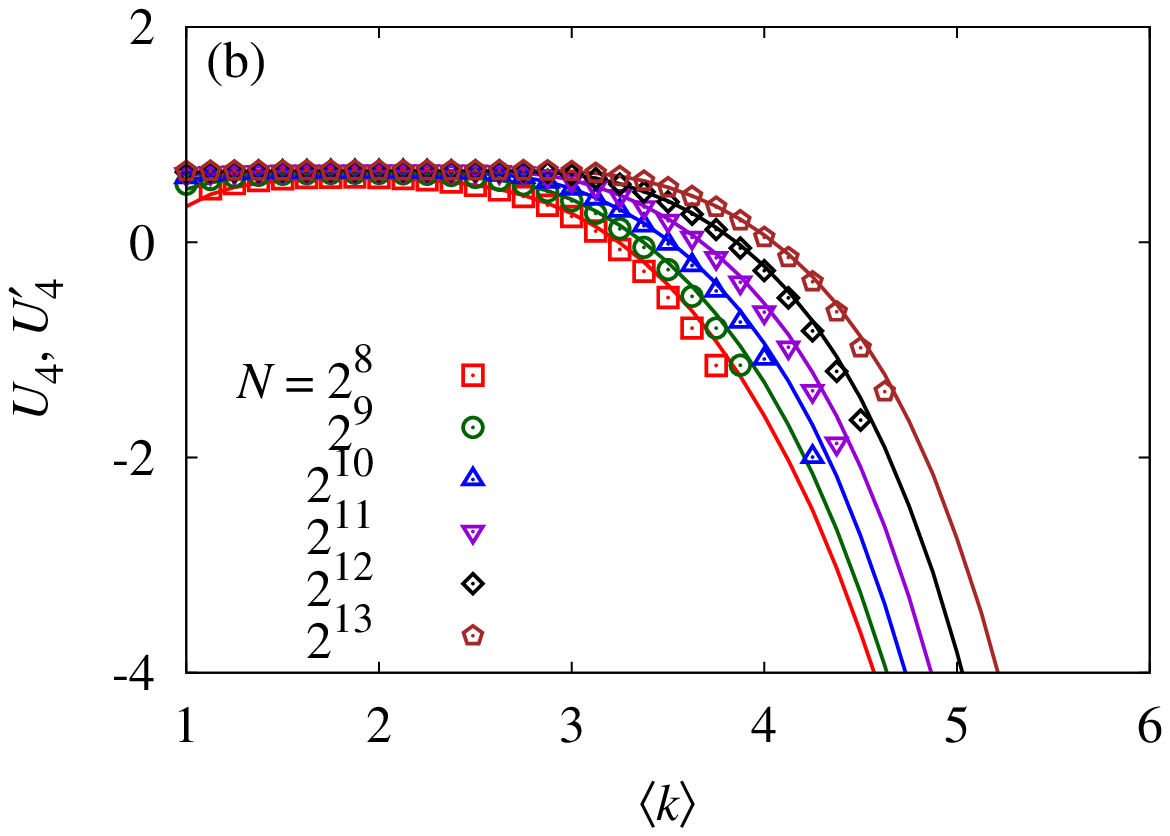}
    \caption{\label{fig:contribution} (Color online)
    Samplet-to-sample fluctuations caused by the contributions of network structure
    and initial spin configuration are measured for $10^2$ initializations of $10^5$ network realizations.
    (a) Variances of $l_A$ measured in two different ways
    (refer to the main text for details) agree well with each other,
    implying that the mean of $l_A$ distribution is unaffected
    by the exact structure of each particular ER network.
    (b) A similar result is observed for Binder cumulants of $l_A$.
    Here open symbols represent $\left(\sigma^2,~ U_4\right)$
    and solid lines represent $\left(\sigma^{\prime 2},~U_4^{\prime}\right)$, respectively.}
\end{figure*}

Based on the results presented so far, we claim that the $l_A$
distribution in the thermodynamic limit can be represented as a
combination of at most two delta peaks. The nature of the left
peak is easily understood: it represents the samples that almost
reach the ground state, with a small (but still finite) fraction
of active links connecting dangling parts of the system. However,
we pose the following question: why should there be another
well-defined peak at a much larger value of $l_A$?

In order to check whether the ER network is the only substrate
with the observed right-peak behavior, we also test zero-temperature
Glauber dynamics in the regular random network where all nodes
have the same degree $k$. The results are shown in Fig.~\ref{fig:rightmean}(a).
It is remarkable that zero-temperature Glauber dynamics in even-degree
regular random networks never produces any ``right peak'' (only a single peak
is formed at or around $l_A = 0$, which may be called the ``left peak''),
which is why the data are not plotted at even values of $\langle k \rangle$.
This seems to be due to large fluctuations provided by blinkers that
can appear only in even-degree nodes. On the other hand, odd-degree
regular random networks produce results very similar to ER networks.
Lack of degree fluctuations in odd-degree regular random networks
yields only small vertical offsets of $[\langle l_A\rangle]_{R}$,
not significantly altering the right-peak properties observed in ER networks.

To get some hints as to whether the location of $[\langle
l_A\rangle]_R$ has any physical meaning, we compare $[\langle
l_A\rangle]_R$ obtained from the previous simulations in ER
networks (whose initial magnetizations are zero, $m_0 = 0$) with
the same quantity obtained from initial states characterized by
the initial active link density $l_0$. For the sake of convenience, let
us denote the latter by a new notation $[\langle l_A\rangle]_R^l$.
Since $l_A$ always decreases with time in Glauber dynamics,
samples initialized by $l_0$ cannot produce right peaks at values
of $l_A$ larger than $l_0$. Therefore, if $[\langle l_A\rangle]_R$
observed in the previous results are truly special, we expect a
drastic difference in the behavior of $[\langle l_A\rangle]_R^l$
between the samples with $l_0 > [\langle l_A\rangle]_R$ and those
with $l_0 < [\langle l_A\rangle]_R$ since the former can produce a
peak around $[\langle l_A\rangle]_R$ while the latter cannot.

Before moving on to the discussion of results, a remark on the preparation of samples with
$l_0$ is in order. Those samples are constructed from zero-magnetization spin configurations
by random flippings of spins that decrease $\left| l_A - l_0 \right|$.
It should be noted that this method does not scan the entire configuration space
of the samples characterized by $l_0$, since the magnetization of each sample
prepared by the method is likely to be very close to zero. However, since the
samples belonging to the right component have magnetization close to zero
in the absorbing states, our initialization method still serves the purpose of
generating samples that are similar to the right-component samples
but different only in the value of $l_A$.

Figure~\ref{fig:rightmean}(b) indeed confirms our previously described expectation: the
samples with $l_0 > [\langle l_A\rangle]_R$ have $[\langle l_A\rangle]_R^l = [\langle l_A\rangle]_R$,
while the right peak disappears in the samples with $l_0 < [\langle l_A \rangle]_R$
(the right peak may still remain for a few more steps of $\langle k \rangle$ as the right peak
produced by samples with $m_0 = 0$ does have a tail to the left of $\left[\langle l_A \rangle\right]_R$,
which is smaller than $l_0$). This indicates that the value of $\left[\langle l_A \rangle\right]_R$
is indeed physically meaningful since it is the only value of active link density
that can be sustained when the magnetization of the system is close to zero.

\subsection{Origin of sample-to-sample fluctuations}

Now we discuss how the fate of each sample is determined,
or what determines whether the system evolves into the left
or the right component of the $l_A$ distribution.
Such sample-to-sample fluctuations come from three factors:
different network structures, different initial configurations,
and the stochasticity of dynamics.

We separate the effect of network structure from that of the other
factors. A natural parameter describing the broadness of the $l_A$
would be its variance, which we may define in the following two ways:
\begin{align*}
    \sigma^2 &= \left[\langle l_A^2 \rangle\right] - \left[\langle l_A \rangle\right]^2, \\
    \sigma'^2 &= \left[\langle l_A^2\rangle -\langle l_A \rangle^2\right].
\end{align*}
Here, $\left[\ldots\right]$ represents an average over different
network realizations, and $\langle\ldots\rangle$ represents an average over
the other factors sharing the common network structure. Therefore,
$\sigma^2$ is the variance of $l_A$ that does not distinguish
between contributions from the three factors mentioned above. On
the other hand, $\sigma'^2$ is the different variance of $l_A$ due
to different initial configuration and the stochasticity of
dynamics, averaged over different network realizations in the
final step of calculation. The difference between $\sigma^2$ and
$\sigma'^2$ can be interpreted as the fluctuations of $l_A$ solely
due to differences in the network structure. Strictly speaking,
$\sigma^2 = \sigma'^2$ implies that $\langle l_A \rangle$ is the
same, irrespective of the exact structure of ER networks as
proven below,
\begin{align*}
\sigma'^2 - \sigma^2 &= \left[\langle l_A \rangle^2\right] - \left[\langle l_A \rangle\right]^2 \\
                  &= \left[\left(\langle l_A \rangle-\left[\langle l_A\rangle\right]\right)^2\right] = 0, \\
\langle l_A \rangle &= \left[\langle l_A \rangle\right].
\end{align*}

Figure~\ref{fig:contribution}(a) shows that $\sigma^2$ and
$\sigma'^2$ are close to each other for various combinations of
$N$ and $\langle k \rangle$. Therefore, the network structure does
not contribute much to the mean values of $l_A$. The same can be
said for the higher moments of $l_A$, such as Binder cumulants
$U_4$ and $U'_4$,
\begin{align*}
    U_4 &= 1-\frac{\left[\langle l_A^4 \rangle\right]}{3\left[\langle l_A^2 \rangle\right]^2}, \\
    U'_4 &= 1-\frac{\left[\langle l_A^4\rangle\right]}{3\left[\langle l_A^2 \rangle^2\right]}
\end{align*}
which are shown in Fig.~\ref{fig:contribution}(b). Hence, we can
conclude that the shape of the $l_A$ distribution is not really
changed by differences in the network structure. Similar results
are observed when we distinguish between the contributions of the
initial configuration and the stochasticity of dynamics. Again, we
can say that the shape of the $l_A$ distribution is not affected by
the difference among initial configurations.

\begin{figure}
    \centering
    \includegraphics[width=0.9\columnwidth]{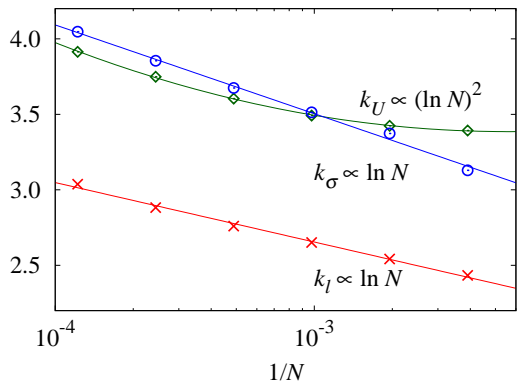}
    \caption{\label{fig:kc}(Color online)
    The $N$-dependence of three specific values of $\langle k
    \rangle$: $k_U~\left(\diamond \right),\ k_{\sigma}\left(\circ\right),\
    k_{l}~\left(\times\right)$. They are possible indicators of the disappearance
    of the left peak (see the main text for their precise definitions).
    }
\end{figure}

\subsection{$P\left(l_A\right)$ in the thermodynamic limit}

Finally, we return to the question posed early in this section: is
the right peak the only lasting peak in the thermodynamic limit?
To answer this question, we define possible indicators of
the point at which the left peak becomes unobservable. The
$N$-dependence of those indicators may reveal whether the left
peak survives in the thermodynamic limit.

The curves shown in Fig.~\ref{fig:contribution} reflect
the $\langle k \rangle$-dependence of $P\left(l_A\right)$.
As $\langle k \rangle$ grows, we observe the shift of
the single (right) peak to the right, and then the formation of the
left peak, and finally the increasing inter-peak separation with the
dominance of the left peak, as Fig.~\ref{fig:lA_dist}(a) illustrates.
This produces the single-peaked curves as shown in Fig.~\ref{fig:contribution}(a).
Although not exhibited here, $\left[\langle l_A \rangle\right]$ shows a similar
behavior. Meanwhile, the Binder cumulant $U_4$ indicates the
separation between two peaks, whose value decreases far below zero
as the separation increases indefinitely with $\langle k \rangle$.

To investigate the finite-size scaling of Fig.~\ref{fig:contribution},
we define the three specific values of $\langle k \rangle$ as
\begin{itemize}
\item{$k_U$: the largest $\langle k \rangle$ that satisfies $U_4=0$,}
\item{$k_{\sigma}$: the location of $\max\sigma^2$,}
\item{$k_{l}$: the location of $\max\left[\langle l_A \rangle\right]$.}
\end{itemize}
Taking those three points as possible indicators of the critical
value of $\langle k\rangle$ at which the left peak is barely observable,
we observe that all the three values increases indefinitely with $N$
(see Fig.~\ref{fig:kc}). This implies that as $N$ goes to infinity,
the $l_A$ distribution is single-peaked at any finite $\langle k \rangle$,
and that the peak is far away (as it is actually the ``right'' peak)
from $l_A = 0$. Hence, this observation again supports our previous
claim that the right peak is actually the only lasting peak in the
thermodynamic limit.

\section{\label{sec:discussion}Summary and Discussions}

In summary, we have studied the absorbing states of
zero-temperature Glauber dynamics in quenched ER-type random
networks with zero-magnetization initial states.
Each absorbing state has been represented by its active link density $l_A$.
We have found that the structure of the $l_A$ distribution consists of two
self-averaging peaks, the left one with samples almost reaching
the ground state and the right one with samples divided by large
stable domain walls. The mean of the left-peak (right-peak) component
seems to be a monotonically decreasing (increasing) function of $\langle k \rangle$.
In particular, we focus on the finite-size scaling
analysis of various physical properties for $P(l_A)$, which
suggests that the fraction of the right-peak component stays finite
in the thermodynamic limit, even if the value of $\langle k \rangle$
is pretty high. We have also gathered circumstantial
evidence that the right peak is actually the only lasting peak in
the thermodynamic limit.

Our results also have some implications on opinion dynamics
studies. Some earlier studies of zero-temperature Glauber dynamics
as an opinion dynamics model have simply assumed that when there
are sufficiently many links, the portion of samples that do not
reach the ground state becomes negligible. Thus, they measured
``consensus time,'' which is the time required for the sample to
reach the ground state in order to obtain the scaling relationship
between the system size and the speed of relaxation. We suggest
that consensus time is not a suitable measure since almost all the
samples fail to reach the ground state in the thermodynamic limit
and increasingly many of them end up far away from the ground
state as $\langle k \rangle$ gets larger~\cite{ODComment}. Data collapse shown in
Fig.~\ref{fig:separation} (the results of \cite{Castellano2005} as
well) is a better numerical approach to such a problem. The final
location of the right peak is very well-defined and it seems to
represent the only value of $l_A$ that yields stable up-down
symmetric configurations. Therefore, interfacial noise that causes
sufficiently large fluctuations in $l_A$ always drives the system
out of absorbing states and helps them relax to the ground state,
as illustrated by the role played by blinkers in even-degree
regular random networks. This leads us to expect that there should
be an analytical argument for the location of the right peak,
which we unfortunately could not formulate. As the right peak
observed in odd-degree regular random networks shows a
qualitatively similar behavior, an intuitive argument for the
nature of the right peak in one substrate may help us understand
the corresponding phenomenon in the other substrate.

Finally, we have shown that the properties of $P(l_A)$ are not
much affected by either the particular network structure or
initial configuration, where both peaks can be reached purely from
the stochasticity of dynamics. This may be due to the
characteristic homogeneity of ER networks, as any randomly
generated ER network in any randomly chosen symmetric initial
configurations is likely to be very similar to another ER network
prepared in the same way. It would be interesting to check whether
the same observation can be made for more topologically heterogeneous
substrates such as scale-free networks. One should note that this
argument for the ``irrelevance of structure and configuration''
should be taken carefully. As Uchida and Shirayama have
shown~\cite{Uchida2007}, the biased choice of initial
configuration does affect the properties of $P(l_A)$. The
contributions from such exceptional cases are negligible in the
unbiased ensemble of samples we have considered in this paper.

\section*{Acknowledgement}

The work was supported by the National Research Foundation of
Korea (NRF) grant funded by the Korean Government (MEST) (No.
2011-0011550) (MH) and partially supported by the NRF grant funded
by the MEST (No. 2011-0028908) (YB, HJ). M.H. would acknowledge
the generous hospitality of KIAS for Associate Member Program,
where the main idea was initiated.


\begin{thebibliography}{99}
\bibitem{ODComment} Note that opinion dynamics models deal with
single-component systems. Our study is about the ER network, which is naturally
composed of multiple components. One should bear this difference
in mind when interpreting our results. Still, we can safely say that
our argument against the concept of consensus time generalizes to
connected systems. At large values of $\langle k \rangle$,
the giant component's contribution to $l_A$ is overwhelmingy large.
Therefore, if the whole network is stuck far away from the ground state,
so is the giant component.
\end{thebibliography}
\end{document}